\documentstyle[12pt]{article}
\def\beq{\begin{equation}}
\def\eeq{\end{equation}}

\def\lap{\lower.5ex\hbox{$\; \buildrel < \over \sim \;$}}
\def\gap{\lower.5ex\hbox{$\; \buildrel > \over \sim \;$}}

\input psfig

\begin{document}
\topskip 2cm 
\begin{titlepage}

\begin{center}
{\large\bf THE STANDARD COSMOLOGICAL MODEL} \\
\vspace{2.5cm}
{\large P. J. E. Peebles} \\
\vspace{.5cm}
{\sl Joseph Henry Laboratories\\ 
Princeton University\\
Princeton, NJ, USA }\\
\vspace{2.5cm}
\vfil
\begin{abstract}
We have a well-established standard model for cosmology and
prospects for considerable additions from work in progress.
I offer a list of elements of the standard model, comments on
controversies in the interpretation of the evidence in
support of this model, and assessments of the directions 
extensions of the standard model seem to be taking. 
\end{abstract}

\end{center}
\end{titlepage}

\section{Introduction}
In the present almost frenetic rate of advance of cosmology it
is useful to be reminded that the big news this year is the
establishment of evidence, by two groups (\cite{Perlmutter},
\cite{Reiss}), of detection of the relativistic curvature of the
redshift-magnitude relation. The measurement was proposed in
the early 1930s. Compare this to the change in the issues in
particle physics since 1930. The slow evolution of cosmology has
allowed ample time for us to lose sight of which elements are
reasonably well established and which have been 
adopted by default, for lack of more reasonable-looking
alternatives. Thus I think it is appropriate to
devote a good part of my assigned space to a discussion of what
might be included in the standard model for cosmology. I then
comment on additions that may come out of work in progress. 

\section{The cosmological model}

Main elements of the model are easily listed: in the
large-scale average the universe is close to homogeneous, and
has expanded in a near homogeneous way from a denser
hotter state when the 3~K cosmic background radiation was
thermalized. 

The standard cosmology assumes conventional physics, including
general relativity theory. This yields a successful account of
the origin of the light elements, at expansion factor 
$z\sim 10^{10}$. Light element formation tests the relativistic
relation between expansion rate and mass density, but this
is not a very searching probe. The 
cosmological tests discussed in \S 3 could considerably improve
the tests of general relativity. 

The model for the light elements seems to require that the
mass density in baryons is less than that needed to account for
the  peculiar motions of the galaxies. It is usually assumed that
the remainder is nonbaryonic (or acts that way). Our reliance on
hypothetical dark matter is an embarrassment; a laboratory
detection would be exceedingly welcome. 

In the past decade many discussions assumed the
Einstein-de~Sitter case, in which there are negligibly small
values for the curvature of sections of 
constant world time and Einstein's cosmological constant
$\Lambda$ (or a 
term in the stress-energy tensor that acts like one).
This is what most of would have chosen if we
were ordering. But the evidence from the relative velocities of
the galaxies has long been that the mass density is less than the 
Einstein-de~Sitter value \cite{Peeb86}, and other more recent
observations, notably the curvature of the redshift-magnitude
relation (\cite{Perlmutter}, \cite{Reiss}), point in the 
same direction. Now there is 
increasing interest in the idea that we live in a universe in
which the dominant term in the 
stress-energy tensor acts like a decaying cosmological
constant (\cite{Ozer} - \cite{Huey}). This is not part of
the standard model, of course, but as discussed in \S 3 the
observations seem to be getting close to useful constraints on
space curvature and $\Lambda$.

We have good reason to think structure formation on the scale of
galaxies and larger was a result of the gravitational growth of
small primeval departures from homogeneity, as described by
general relativity in linear perturbation theory. The
adiabatic cold dark matter (ACDM) model gives a fairly definite
and strikingly successful prescription for the initial conditions 
for this gravitational instability picture, 
and the ACDM model accordingly is widely used in analyses of
structure formation. But we
cannot count it as part of the standard model because there is at
least one viable alternative, the isocurvature model mentioned in
\S 3.3. Observations in progress likely will eliminate at least
one, perhaps establish the other as a good
approximation to how the galaxies formed, or perhaps lead us to
something better.

The observational basis for this stripped-down standard  
model is reviewed in references \cite{Peeb93} and \cite{Peeb91}.
Here I comment on some issues now under discussion.  

\subsection{The cosmological principle}
Pietronero \cite{Piet} argues that the evidence from redshift
catalogs and deep galaxy counts is that the galaxy distribution
is best described as a 
scale-invariant fractal with dimension $D\sim 2$. Others
disagree (\cite{Davis}, \cite{Scaramella}). I am
heavily influenced by another line of argument:
it is difficult to reconcile a fractal universe with the
isotropy observed in deep surveys (examples of which are
illustrated in Figs. 3.7 to 3.11 in \cite{Peeb93} and are
discussed in connection with  the fractal universe in pp. 209 -
224 in \cite{Peeb93}). 

\begin{figure}
\centerline{\psfig{file=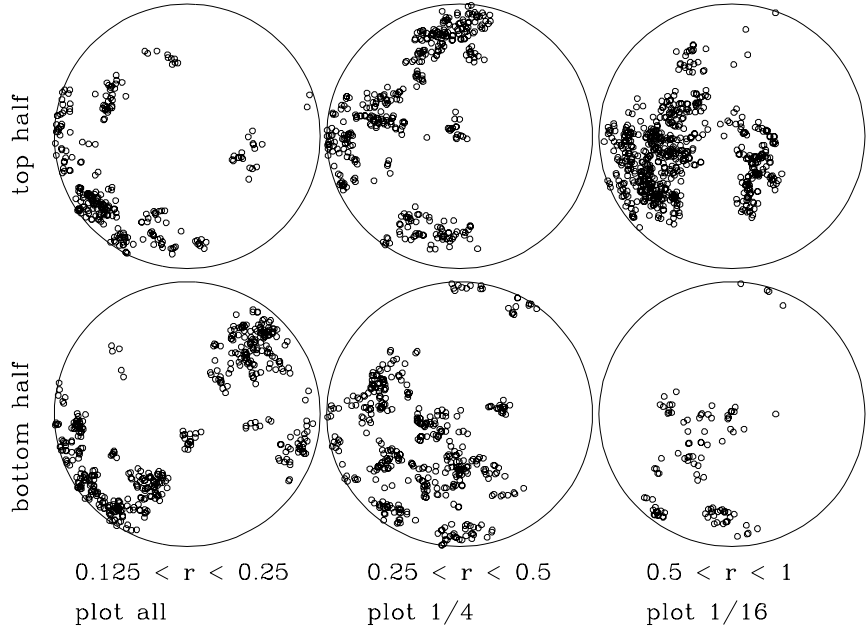,width=4.5truein,clip=}}
\caption{Angular distributions of particles in a realization of a
fractal with dimension $D=2$ viewed from one of the particles in
the realization. The fraction of particles plotted in each
distance bin has been scaled so the expected number of particles 
plotted is the same in each bin.}
\vskip -1.0truecm
\end{figure}

Fig.~1 shows angular positions of particles in three ranges of
distance from a particle in a fractal realization with dimension
$D=2$ in 
three dimensions. At $D=2$ the expected number of neighbors 
scales with distance $R$ as $N(<R)\propto R^2$, and 
I have scaled the fraction of particles plotted as $R^{-2}$ to
get about the same number in each plot. 
The fractal is constructed by placing a stick of length
$L$, placing on either end the centers of sticks of length
$L/\lambda$, where $\lambda = 2^{1/D}$, with random orientation,
and iterating to smaller and larger scales. The  
particles are placed on the ends of the
shortest sticks in the clustering hierarchy. This construction
with $D=1.23$ (and some adjustments to fit the galaxy three- and
four-point correlation functions) gives a good description of the
small-scale galaxy clustering \cite{SP}. The fractal in
Fig.~1, with $D=2$, the dimension
Pietronero proposes, does not look at all like deep sky maps of
galaxy distributions, which show an approach to isotropy with
increasing depth. This cannot happen in a scale-invariant
fractal: it has no characteristic length. 

A characteristic clustering length for galaxies may be expressed in
terms of the dimensionless two-point correlation function
defined by the joint probability of finding galaxies centered in
the volume elements $dV_1$ and $dV_2$ at separation $r$, 
\beq
	dP = n^2[1+\xi _{gg}(r)]dV_1dV_2.
\label{eq:xigg}
\eeq
The galaxy two-point function is quite close to a power law,
\beq
	\xi = (r_o/r)^\gamma,\quad\gamma = 1.77,\quad 
	10\hbox{ kpc}\lap hr\lap 10\hbox{ Mpc},
\label{eq:xiggparameters}
\eeq
where the clustering length is 
\beq
	r_o=4.5\pm 0.5\hbox{ Mpc},
\eeq
and the Hubble parameter is 
\beq
	H_o=100h\hbox{ km s}^{-1}\hbox{ Mpc}^{-1}.
\eeq
The rms fluctuation in galaxy counts in a randomly placed sphere
is $\delta N/N = 1$ at sphere radius $r=1.4r_o\sim 6h^{-1}$~Mpc, to
be compared to the Hubble distance (at which the recession
velocity approaches the velocity of light),
$cH_o^{-1}=3000h^{-1}$ Mpc. 

The isotropy observed in deep sky maps is consistent with a
universe that is inhomogeneous  
but spherically symmetric about our position. There are
tests, as discussed by Paczy\'nski and Piran \cite{Bohdan}. For
example, we have a successful theory for the 
origin of the light elements as remnants of the expansion and
cooling of the universe through $kT\sim 1$ MeV \cite{Schramm}.
If there were a strong radial matter density gradient out to the
Hubble length we could be using the wrong local entropy per
baryon, based on conditions at the Hubble length where
the CBR came from, yet the theory seems to be successful. But to
most people the compelling argument is that distant galaxies look
like equally good homes for observers like us: it would be
startling if we lived in one of the very few close to the center 
of symmetry.

Mandelbrot \cite{Benoit} points out that other fractal
constructions could do better than the one in Fig. 1. His example
does have  more particles in the voids defined by the strongest
concentrations in the sky, but it seems to me to share the
distinctly clumpy character of Fig.~1. It would be interesting to  
see a statistical test. A common one expands the angular
distribution in a given range of distances in spherical
harmonics,   
\beq
	a_l^m = \int d\Omega\,\sigma (\Omega )\,Y_l^m(\Omega ),
\label{eq:alm}
\eeq
where $\sigma$ is the surface mass density as a function of
direction $\Omega$ in the sky. The integral becomes
a sum if the fractal is represented as a set of particles.
A measure of the angular fluctuations is
\beq
	e_l = l\sum _{-l<m<l} |a_l^m|^2/(a_0^0)^2, 
\eeq
where 
\beq
	\langle\sigma ^2\rangle /\langle\sigma\rangle ^2 - 1
	=\sum _{l \geq 1} e_l/l.
\label{eq:variance}
\eeq
In the approximation of the sum as an integral $e_l$ is
the contribution to the variance of the angular distribution per
logarithmic interval of 
$l$. It will be recalled that the zeros of the real and
imaginary parts of $Y_l^m$ are at separation $\theta =\pi /l$ in
the shorter direction, except where the zeros crowd together near
the poles and $Y_l^m$ is close to zero. Thus $e_l$ is the
variance of the fractional fluctuation in density across the sky
on the angular scale $\theta\sim\pi /l$ and in the chosen range of
distances from the observer.

I can think of two ways to define the dimension of a fractal that 
produces a close to isotropic sky. First, each octant of a full
sky sample has half the diameter of the full sample, so 
one might define $D$ by the fractional departure of 
the mean density within each octant from the mean in the full
sample,  
\beq
	(e_2)^{1/2}\sim 2^{3-D} - 1.\label{eq:e_2}
\eeq
Thus in Fig. 1, with $D=2$, the quadrupole anisotropy $e_2$ is on
the order of unity. Second, one can use the idea that the mean
particle density varies with distance $r$ from a particle as
$r^{-(3-D)}$. Then the small angle (large $l$) Limber
approximation to the angular correlation function $w(\theta )$ is
\cite{LSS} 
\beq
	1+w(\theta =\pi /l)\sim \int ^l e_l\, dl/l \propto
	\int _0^1du [u^2 + (\pi /l)^2]^{-(3-D)/2}.
\label{eq:e_l}
\eeq
To find $e_l$ differentiate with respect to $l$. 
At $D=2$ this gives $e_l\sim 1$: the surface density fluctuations
are independent of scale. At $0<3-D\ll 1$, $e_l\sim (3-D)/l$. 
The X-ray background fluctuates by about $\delta f/f\sim 0.05$ at 
$\theta =5^\circ$, or $l\sim 30$. This is equivalent to 
$D \sim 3 - l(\delta f/f)^2\sim 2.9$ in the fractal model in
Eq.~(\ref{eq:e_l}).

The universe is not exactly homogeneous, but it seems to be
remarkably close to it on the scale of the Hubble length. 
It would be interesting to know whether there is a fractal
construction that allows a significantly larger 
value of $3-D$ for given $e_l$ than in this calculation. 

\subsection{The Hubble redshift-distance relation}
Expansion that preserves homogeneity requires that the mean rate of
change of separation of pairs of galaxies with separation $R$
varies as the Hubble law, 
\beq
	v = HR.\label{eq:hl}
\eeq
The redshift-distance relation for type Ia supernovae gives an
elegant demonstration of this relation (\cite{Perlmutter},
\cite{Reiss}). Arp (\cite{Arp}, \cite{Arp2})
points out that such precision tests do not directly apply to the
quasars, and he finds fascinating evidence in sky maps for
associations of quasars with galaxies at distinctly lower
redshifts. But there is a counterargument, along lines pioneered
by Bergeron \cite{Bergeron}, as follows.

A quasar spectrum may contain absorption lines characteristic of
a cloud of neutral atomic hydrogen at surface density 
$\Sigma _{\rm HI}\gap 3\times 10^{17}$ atoms~cm$^{-2}$. If this
absorption system is at redshift $z\lap 1$ a galaxy at the
same redshift is close enough that there is a reasonable chance 
observing it, and with high probability an optical image does
show a galaxy close to the quasar and at the redshift of the
absorption lines (\cite{Steidel},
\cite{Lanzetta}). Also, when a galaxy image 
appears in the sky close to a quasar at higher redshift then with
high probability the quasar spectrum has absorption lines at the
redshift of the galaxy. We have good evidence the galaxy is
at the distance 
indicated by its redshift. We can be sure the quasar is behind
the galaxy: the quasar light had to have passed through the
galaxy to have produced the absorption lines. If quasars were not
at their cosmological distances we 
ought to have examples of a quasar appearing close to the
line of sight to a lower redshift galaxy and without the
characteristic absorption lines produced by the gas in and around
the galaxy.  

Arp's approach to this issue is important, but I am influenced by
what seems to be this direct and clear interpretation of the
Bergeron effect, that indicates redshift is a good measure of
distance for quasars as well as galaxies. 

\subsection{The expansion of the universe}

In the relativistic Friedmann-Lema\^\i tre cosmological model the
wavelength of a freely propagating photon is stretched in
proportion to the expansion factor from the epoch of emission to
detection:
\beq
	1+z={\lambda _{\rm obs}\over\lambda _{\rm em}} =
	{a_{\rm obs}\over a_{\rm em}}.
\label{eq:redshift}
\eeq
The first expression defines the redshift $z$ in terms of the
ratio of observed wavelength to wavelength at emission. The
cosmological expansion parameter $a(t)$ is proportional to the
mean distance between conserved particles.

The most direct evidence that the redshift is a
result of expansion is the thermal spectrum of the 
CBR \cite{Fixsen}. In a tired light model in a static
universe the photons suffer a redshift that is proportional to 
the distance travelled, but in the absence of absorption or
emission the photon number density remains constant. In this case a
significant redshift makes an initially thermal spectrum distinctly 
not thermal and inconsistent with the measured CBR spectrum. 
One could avoid this by assuming the mean free path for 
absorption and emission of CBR photons is much shorter than the
Hubble length, so relaxation to thermal equilibrium is much
faster than the rate of distortion of the spectrum by the
redshift. But this opaque universe is quite inconsistent 
with the observation of radio galaxies at redshifts $z\sim 3$ at
CBR wavelengths. That is, the universe cannot have an optical
depth large enough to preserve a thermal CBR spectrum in a tired
light model. In the standard world model the expansion has two
effects: it redshifts the photons, as $\lambda\propto a(t)$, and
it dilutes the photon number density, as $n\propto a(t)^{-3}$.  
The result is to cool the CBR while keeping its spectrum thermal. 
Thus the expanding universe allows a self-consistent picture: the
CBR was thermalized in the past, at a time when when the universe
was denser, hotter, and optically thick.

I have not encountered any serious objection to this argument;
the issue is the expansion factor. In the relativistic
Friedmann-Lema\^\i tre model the expansion of the universe traces 
back at least as far as redshift $z\sim 10^{10}$, when the light
elements formed in observationally reasonable amounts \cite{Schramm}. 
In the model of Arp {\it et al.} \cite{Arp3} the expansion
and cooling traces back to a redshift only moderately 
greater than the largest observed values, $z\sim 5$, when there
would have been a burst of creation of matter and radiation
followed by rapid clearing of the dust that thermalized the
radiation. The Arp {\it et al.} picture
for the origin of the light elements has not been
widely debated. If it were agreed that it is viable then a
choice between this and the Friedmann-Lema\^\i tre model would
depend on other tests, such as the angular fluctuations in the
CBR, as discussed next.

\begin{figure}
\centerline{\psfig{file=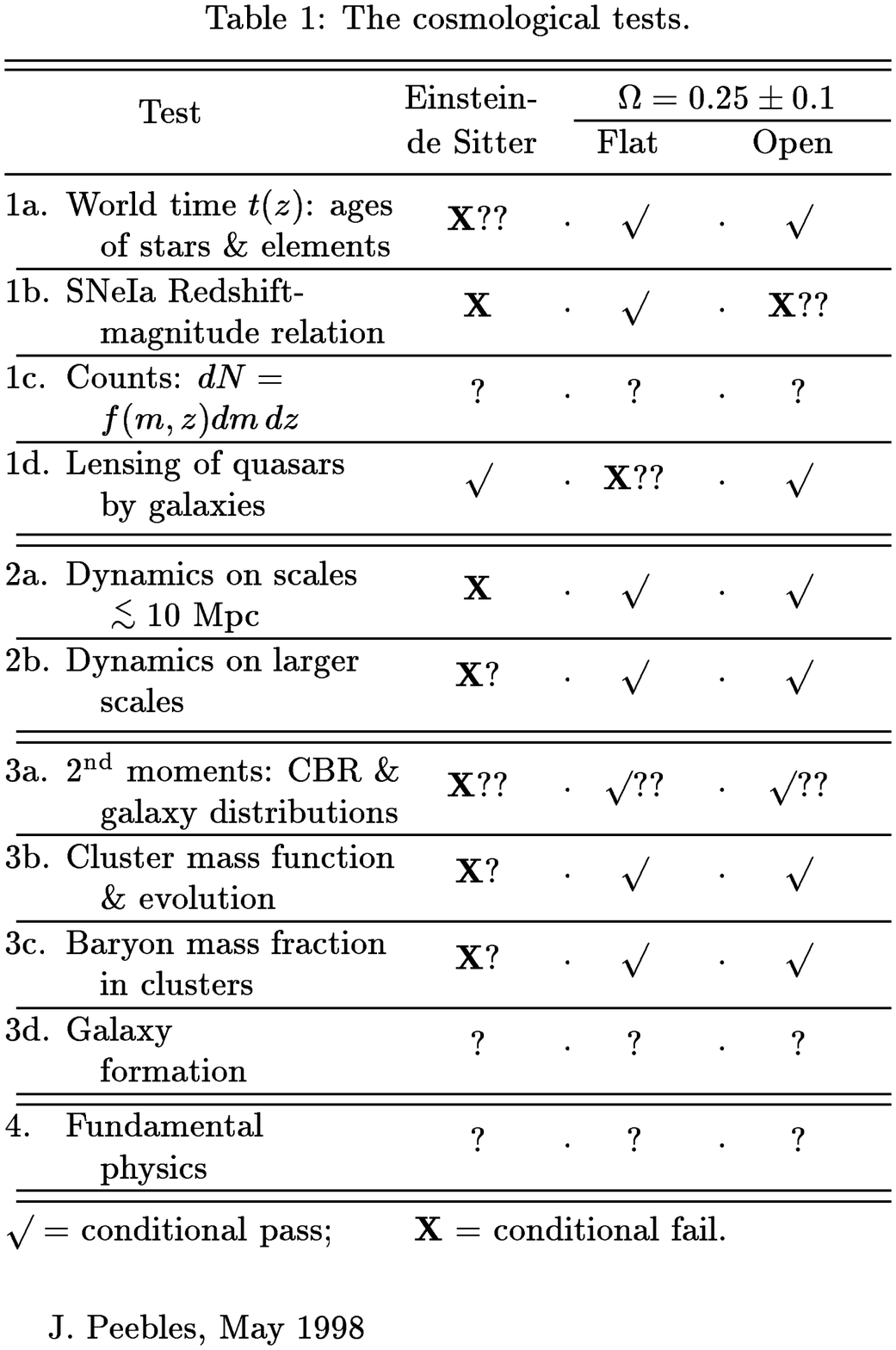,width=4.1truein,clip=}}
\vskip -1.0truecm
\end{figure}

\section{The cosmological tests}

The tests in Table 1 are organized in four
categories: spacetime geometry, galaxy peculiar velocities, 
structure formation, and early universe physics. I offer
grades for three sets of parameter choices. As the tests improve
we may learn that one narrowly constrained set of 
values of the cosmological parameters receives consistent passing 
grades, or else that we have to cast our theoretical net more
broadly.  

\subsection{Spacetime geometry}

In the relativistic Friedmann-Lema\^\i tre cosmological model 
the mean spacetime geometry (ignoring curvature fluctuations
produced by local mass concentrations in galaxies and systems of
galaxies) may be represented by the line element
\beq
	ds^2 = dt^2 - a(t)^2\left[ {dr^2\over 1\pm r^2/R^2}
	+ r^2(d\theta ^2 + \sin ^2\theta d\phi ^2)\right] , 
\label{eq:lineelement}
\eeq
where the expansion rate satisfies the equation
\beq
	H^2 = \left( \dot a\over a\right) ^2 
	= {8\over 3}\pi G\rho \pm {1\over a^2R^2} + 
	{\Lambda\over 3},
\eeq
which might be approximated as
\beq
	H^2= H_o^2[\Omega (1+z)^3 + \kappa (1+z)^2 +\lambda ].
\label{eq:cos_pars}
\eeq
The last equation defines the fractional contributions to the
square of the present Hubble parameter $H_o$ by matter, space
curvature, and the 
cosmological constant (or a term in the stress-energy tensor that
acts like one). The time-dependence assumes pressureless matter
and constant $\Lambda$. Other notations are in the literature;  
a common practice in the particle physics community to add
the matter and $\Lambda$ terms in a new density parameter, 
$\Omega ' = \Omega + \lambda$. I prefer keeping them separate, 
because the observational signatures of $\Omega$ and $\lambda$
can be quite different.

By 1930 people understood how one would test the space-time
geometry in these equations, and as I mentioned there is at last
direct evidence for the detection of one of the effects, the 
curvature of the relation between redshift and apparent magnitude
(\cite{Perlmutter}, \cite{Reiss}). As indicated in line 1b, the
measured curvature is  
inconsistent with the Einstein-de~Sitter model in which $\Omega =1$
and $\lambda = 0 =\kappa$. The measurements also disagree with a
low density model with $\lambda =0$, though the
size of the discrepancy approaches the size of the error flags,
so I assign a weaker failing grade for this case. 
The measurements are magnificent. The 
issue yet to be thoroughly debated is whether
the type Ia supernovae observed at redshifts $0.5\lap z\lap 1$
are drawn from essentially the same population as the nearer ones.

In a previous volume in this series Krauss \cite{Krauss}
discusses the time-scale issue. Stellar
evolution ages and radioactive decay ages do not rule out the
Einstein-de~Sitter model, within the still considerable 
uncertainties in the measurements, but the longer expansion time
scales of the low $\Omega$ models certainly relieve the problem of
interpretation of the measurements. Thus I enter a
tentative negative grade for the Einstein-de~Sitter model in line
1a. 

In the analysis by Falco {\it et al.} \cite{lensing} of the rate
of lensing of quasars by foreground galaxies (line~1d) for a
combined sample of lensing events detected in the optical and
radio, the $2\sigma$ bound on the density parameter in a
cosmologically flat ($\kappa = 0$) universe 
is $\Omega > 0.38$. The SNeIa redshift-magnitude
relation seems best fit by $\Omega = 0.25$, $\lambda = 0.75$, a
possibly significant discrepancy. A serious uncertainty in the
analysis of the lensing rate is the number density of early-type
galaxies in the high surface density branch of the fundamental
plane at luminosities $L\sim L_\ast$, the luminosity of the
Milky Way. If further tests confirm an inconsistency of the
lensing rate and the redshift-magnitude relation the 
lesson may be that $\lambda$ is dynamical, rolling to zero, as
Ratra \&\ Quillen \cite{Ratra} point out.  

\subsection{Biasing and large-scale velocities}

The relation between the mass density parameter $\Omega$ and the 
gravitational motions of the galaxies is an issue
rich enough for a separate category in Table 1. It has been
known for the past decade that if galaxies were fair tracer of
mass then the small-scale relative
velocities of the galaxies would imply that $\Omega$ is well
below unity \cite{Peeb86}. If the mass distribution were 
smoother than that of the galaxies, the smaller mass
fluctuations would require a 
larger mean mass density to gravitationally produce the observed
galaxy velocities. Davis, Efstathiou, Frenk \&\ White \cite{DEFW}
were the first to show that such a biased distribution of galaxies 
relative to mass readily follows in numerical N-body
simulations of the growth of structure, and the demonstration has
been repeated in considerable detail (\cite{bias1},
\cite{bias2}, and references therein). This is a serious argument
for the biasing effect. But here are three arguments for
the proposition that galaxies are fair tracers of mass for the
purpose of estimating $\Omega$. 

First, in many numerical simulations dwarf galaxies are less
strongly clustered than giants. This is reasonable, for if much
of the mass were in the voids defined by the giant galaxies, as
required if $\Omega = 1$, then surely there would be remnants of
the suppressed galaxy formation in the voids, irregular galaxies
that bear the stigmata of a hostile early environment. The first
systematic redshift survey showed that the distributions of low
and high luminosity galaxies are strikingly similar \cite{CfA}. No
survey since, in 21-cm, infrared, ultraviolet, or low surface
brightness optical, has revealed a void population. There is a 
straightforward interpretation: the voids are nearly
empty because they contain little mass.

Second, one can use the galaxy two-point correlation function in
Eq.~(\ref{eq:xigg}) and the mass autocorrelation function 
$\xi _{\rho\rho }$ from a numerical simulation of  structure
formation to define the bias function
\beq
  b(r,t) = \left[ \xi _{gg}(r,t)/\xi _{\rho\rho}(r,t)\right] ^{1/2}.
\label{eq:bias}
\eeq
In numerical simulations $b$ typically
varies quite significantly with separation and redshift 
\cite{bias1}. That is, the galaxies give a biased
representation of the statistical character of the mass
distribution in a typical numerical simulation. The issue 
is whether the galaxies, or the models, or both, are biased
representations of the statistical character of the real mass
distribution. What particularly strikes me is the observation
that the low order galaxy correlation functions have some simple
properties. The galaxy two-point function is close to a power
law over some three orders of magnitude in separation 
(Eq. \ref{eq:xiggparameters}). The value of the power law index
$\gamma$ changes little back to redshift $z\sim 1$. Within the
clustering length $r_o$ the higher order correlation functions
are consistent with a power law fractal. A reasonable presumption
is that the regularity exhibited by the galaxies reflects a like
regularity in the mass, because galaxies trace mass. I am
impressed by the power of the numerical simulations, and believe
they reflect important aspects of reality, but do not think we
should be surprised if they do not fully represent other aspects,
such as relatively fine details of the mass distribution.

\begin{figure}
\centerline{\psfig{file=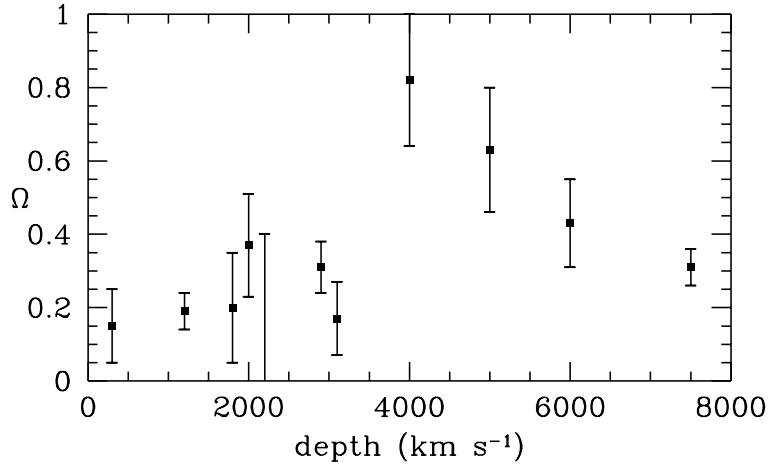,width=4.truein,clip=}}
\caption{The density parameter derived from galaxy peculiar
motions on the assumption galaxies trace mass. From the left, the
estimates are based on the Local Group of galaxies, clusters of
galaxies, the peculiar infall toward the Virgo Cluster
\cite{Tonry}, and the analyses in references \cite{v1} to
\cite{v8}.}  \vskip -1.0truecm
\end{figure}

The third argument deals with the idea that blast waves or
radiation from the formation of a galaxy may have affected the
formation of nearby galaxies, producing scale-dependent bias. In
this case the apparent value of the density parameter derived  
from gravitational motions within systems of galaxies on the
assumption galaxies trace mass would be 
expected to vary with increasing scale, approaching the true
value when derived from relative motions on scales larger than
the range of influence of a forming galaxy. Fig.~2 shows a test.
The abscissa at the entry for clusters of galaxies is the 
comoving radius of a sphere that contains the mass within the
Abell radius. The estimates at larger scales are plotted at
approximate values of the radius of the sample. If it were not
for the last two points at the right-hand side of Fig.~2, 
one might conclude that the apparent density parameter is
increasing to the true value $\Omega\sim 1$ at
$R\sim 50h^{-1}$~Mpc. But considering the last two points,
and the sizes of the error flags, it is difficult to
see any evidence for scale-dependent bias. 

I assign a strongly negative grade for the Einstein-de~Sitter
model in line 2a in Table~1, based
on galaxy motions on relatively small scales, because biasing
certainly is required if $\Omega =1$ and I have argued there
is no evidence for it. The more tentative grade in line 2b is
based on Fig.~2: the apparent value of the density parameter does
not seem to scale with depth.  

\subsection{Structure formation}

The Friedmann-Lema\^\i tre model is unstable
to the gravitational growth of departures from a homogeneous mass
distribution. The present large-scale homogeneity could have
grown out of primeval chaos, but the initial conditions would
be absurdly special. That is, the Friedmann-Lema\^\i tre model
requires that the present structure---the clustering of mass in
galaxies and systems of galaxies---grew  
out of small primeval departures from homogeneity. The
consistency test for an acceptable set of cosmological parameters
is that one has to be able to assign a physically sensible
initial condition that evolves into the present structure of the
universe. The constraint from this consideration in 
line~3c is discussed by White {\it et al.} \cite{BaryonC}, and in
line~3b by Bahcall {\it et al.} (\cite{Neta1}, \cite{Neta2}).
Here I explain the cautious ratings in line~3a.  

As has been widely discussed, it may be possible to read the
values of $\Omega$ and other cosmological parameters from the
spectrum of angular fluctuations of the CBR (\cite{Stein} and
references therein). This assumes Nature has kept the evolution
of the early universe simple, however, and we have hit on
the right picture for its evolution. We may know in the next few
years. If the precision measurements of the CBR anisotropy from
the MAP and PLANCK satellites match in all detail the prediction
of one of the structure formation models now under discussion it
will compel acceptance. But meanwhile we should bear in mind
the possibility that Nature was not kind enough to have presented 
us with a simple problem. 

\begin{figure}
\centerline{\psfig{file=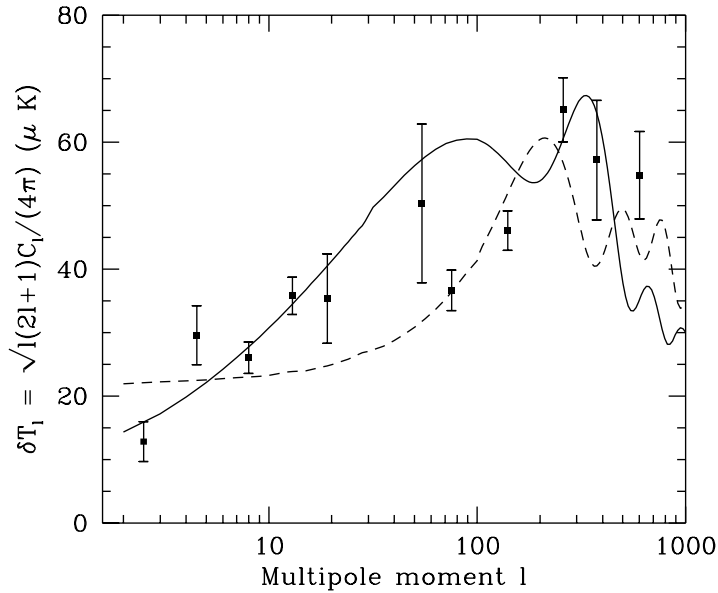,width=3.75truein,clip=}}
\caption{Angular fluctuations of the CBR in low density
cosmologically flat 
adiabatic (dashed line) and isocurvature (solid line) CDM models
for structure formation. The variance of 
the CBR temperature anisotropy per logarithmic interval of angular
scale $\theta =\pi /l$ is $(T_l)^2$, as in Eqs.~(\ref{eq:alm})
to~(\ref{eq:variance}).  Data are from
the compilation by Ratra \cite{RatraComp}.} 
\vskip -1.0truecm
\end{figure}

An example of the possible ambiguity in the interpretation of the
present anisotropy measurements is shown in Fig.~3. The two models
assume the same dynamical actors---cold dark matter (CDM),
baryons, three families of massless neutrinos, and the CBR---but
different initial conditions. In the adiabatic model the primeval
entropy per conserved particle number is homogeneous, the space
distribution of the primeval mass density fluctuations is a
stationary random process with the scale-invariant
spectrum $P(k)\propto k$, and the cosmological parameters are $\Omega =
0.35$, $\lambda = 0.65$, and $h=0.625$ (following \cite{OS}).
The isocurvature initial condition in the other model is that the  
primeval mass distribution is homogeneous---there are no
curvature fluctuations---and structure formation is seeded by an
inhomogeneous composition. In the model  
shown here the primeval entropy per baryon is homogeneous, to
agree with the standard model for light element production, and
the primeval distribution of the CDM has fluctuation spectrum   
\beq
	P(k)\propto k^m, \qquad m = -1.8.
\label{eq:Piso}
\eeq
The cosmological parameters are $\Omega =0.2$, $\lambda =0.8$, and 
$h=0.7$. The lower density parameter produces a more
reasonable-looking cluster mass function for the 
isocurvature initial condition \cite{PeebII}. 
In both models the density parameter in baryons is 
$\Omega _{\rm B}=0.03$, the rest of $\Omega$ is in CDM, and space
sections are flat ($\lambda = 1 -\Omega$).  
Both models are normalized to the large-scale
galaxy distribution. The adiabatic initial condition follows
naturally from inflation, 
as a remnant of the squeezed field that drove the rapid
expansion. A model for the isocurvature condition assumes the CDM
is (or is the remnant of) a massive scalar field that was in the
ground level during inflation and became squeezed to a classical
realization. In the simplest models for inflation this produces 
$m=-3$ in Eq.~(\ref{eq:Piso}). The tilt to $m=-1.8$ requires only
modest theoretical ingenuity \cite{PeebI}. That is, both models
have pedigrees from commonly discussed early universe physics.  

The lesson from Fig.~3 is that at least two families
of models, with different relations between $\Omega$ and the
value of $l$ at the peak, come close to the measurements of the
CBR fluctuation spectrum, within the still substantial
uncertainties. An estimate of $\Omega$ from the CBR 
anisotropy measurements thus may depend on the choice of the
model for structure formation. Programs of measurement of 
$\delta T_l$ in progress should be capable of  
distinguishing between the adiabatic and isocurvature models,
even given the freedom to adjust the shape of $P(k)$.
The interesting possibility is that
some other model for structure formation with a very different
value of $\Omega$ may give an even better fit to the improved
measurements.

I assign a failing grade to the Einstein-de~Sitter model in line
3a because the adiabatic and isocurvature models both prefer low
$\Omega$ (\cite{Gawiser}, \cite{Bond}). I add question marks
to indicate this still is a model-dependent result. 

\subsection{Constraints from fundamental physics}

In their version of Table 1 Dekel, Burstein, \&\ White \cite{DBW}
give the Einstein-de Sitter model the highest grade on
theoretical grounds, and a cosmologically flat model with
$\Lambda$ the next highest grade. The point is well
taken: this is the order most of us would choose. The issue is
whether Nature agrees with our ideas of elegance, or maybe 
prefers physics that produces an open universe (\cite{Gott} -
\cite{Hawking}). Full closure of cosmology may come with the
discovery of physics that predicts the values of $\lambda$ 
and space curvature (Eq.~[\ref{eq:cos_pars}]) in terms of 
the expansion age of the universe, consistent with all the other 
constraints in Table~1. But since we seem to be far from that
goal I am inclined to omit entries in line~4.

\section{Concluding remarks}

We have a secure if still schematic standard model for
cosmology, and the prospect for considerable enlargement
from the application of the cosmological tests. The theoretical
basis for the tests was discovered seven decades ago. 
A significant application likely will take a lot less than seven
more decades: the constraints in Table~1 already are serious, if
debatable, and people know how to do better.  

Application of the tests could yield a set of tightly constrained
values of the cosmological parameters and a clear
characterization of the primeval departure from homogeneity. If
so cosmology could divide at a fixed point, the situation at
$z=10^{15}$, say,  when the universe is well described by a
slightly perturbed Friedmann-Lema\^\i tre model. One branch of
research would analyze evolution from these initial conditions
to the present complex structure of the
universe. The other would search for the physics of the very
early universe that produced these initial conditions. But before
making any long-term 
plans based on this  scenario I would wait to see whether the
evidence really is that the early universe is simple enough to
allow such a division of labor. 

\vskip 1truecm\noindent I am grateful to the organizers for
the invitation to this stimulating meeting. The work
in this paper was supported in part by the USA National Science
Foundation.

\end{document}